\documentstyle[amssymb,prl,aps,epsfig]{revtex}

\begin{document}

\title{Evidence for charged critical fluctuations in underdoped YBa$_{2}$Cu$%
_{3}$O$_{7-\delta }$}
\author{T. Schneider$^{\text{1}}$, R. Khasanov$^{\text{1,2,3}}$, K. Conder$^{\text{2}}$, E. Pomjakushina$^{\text{2}}$, R.~Bruetsch$^{\text{4}}$ and H. Keller$^{\text{1}}$}
\address{$^{\text{(1)}}$ Physik-Institut der Universit\"{a}t Z\"{u}rich,
Winterthurerstrasse 190, CH-8057,Switzerland\\ $^{\text{(2)}}$
Laboratory for Neutron Scattering, ETH Z\"{u}rich and PSI
Villigen, CH-5232 Villigen PSI, Switzerland\\$^{\text{(3)}}$DPMC,
Universit\'e de Gen\`eve, 24 Quai Ernest-Ansermet, 1211 Gen\`eve
4, Switzerland\\ $^{\text{(4)}}$ Laboratory for Material Behavior,
PSI Villigen, CH-5232 Villigen PSI, Switzerland}
\maketitle

\begin{abstract}
We report and analyze in-plane penetration depth measurements in
YBa$_{2}$Cu$_{3}$O$_{7-\delta }$ taken close to the critical
temperature $T_{c}$. In underdoped YBa$_{2}$Cu$_{3}$O$_{6.59}$ we
find consistent evidence for charged critical behavior. Noting
that the effective dimensionless charge $\widetilde{e}=\xi
/\lambda =1/\kappa $ scales as $T_{c}^{-1/2}$, this new critical
behavior should be generically observable in suitably underdoped
cuprates.
\end{abstract}

\bigskip

Close to the critical temperature $T_{c}$ of the
normal-superconductor transition, in a regime roughly given by the
Ginzburg criterion\cite {ffh,tsda,tshkws,book}, order parameter
fluctuations dominate critical properties. In recent years, the
effect of the charge of the superconducting order parameter in
this regime in three dimensions has been studied extensively\cite
{coleman,halperin,dasgupta,folk,kleinert,herbut,herbut2,olsson,folk2,hove,mo}.
While for strong type-I materials, the coupling of the order
parameter to transverse gauge field fluctuations is expected to
render the transition first order \cite{halperin}, it is
well-established that strong type-II materials should exhibit a
continuous phase transition, and that sufficiently close to
$T_{c}$, the charge of the order parameter is relevant
\cite{folk,kleinert,herbut,herbut2,olsson,folk2,hove,mo}. However,
in cuprate superconductors within the fluctuation dominated
regime, the region close to $T_{c}$, where the system crosses over
to the regime of charged fluctuations turns out to be too narrow
to access. For instance, optimally doped
YBa$_{2}$Cu$_{3}$O$_{7-\delta }$, while possessing an extended
regime of critical fluctuations, is too strongly type-II to
observe charged critical
fluctuations\cite{ffh,tsda,tshkws,book,kamal}. Indeed, the
effective dimensionless charge $\widetilde{e}=\xi /\lambda
=1/\kappa $ is in strongly type II superconductors ($\kappa >>1$)
small. The crossover upon approaching $T_{c}$ is thus initially to
the critical regime of a weakly charged superfluid where the
fluctuations of the order parameter are essentially those of an
uncharged superfluid or XY-model \cite{ffh}. Furthermore, there is
the inhomogeneity induced finite size effect which renders the
asymptotic critical regime unattainable\cite{tsbled,tsdc}.
However, underdoped cuprates could open a window onto this new
regime because $\kappa $ is expected to become rather small. Here
the cuprates undergo a quantum superconductor to insulator
transition in the underdoped limit\cite
{book,klosters,tshk,tsphysB,parks} and correspond to a 2D
disordered bosonic system with long-range coulomb interactions.
Close to this quantum transition $T_{c}$, $\lambda _{ab}$ and $\xi
_{ab}$ scale as $T_{c}\propto \lambda _{ab}^{-2}\propto \xi
^{-z}$\cite{book,klosters,tshk,tsphysB,parks}, yielding with the
dynamic critical exponent $z=1$\cite
{book,parks,mpafisher,ca,herbutz1}, $\kappa _{ab}\propto
\sqrt{T_{c}}$. Noting that $T_{c}$ decreases by approaching the
underdoped limit, sufficiently homogeneous and underdoped cuprates
appear to be potential candidates to observe charged critical
behavior.

Here we report and analyze in-plane penetration depth measurements
of underdoped YBa$_{2}$Cu$_{3}$O$_{7-\delta }$ to explore the
evidence for this new critical behavior.
YBa$_{2}$Cu$_{3}$O$_{7-\delta }$ samples were synthesized by
solid-state reactions using high-purity Y$_{2}$O$_{3}$, BaCO$_{3}$
and CuO. The samples were calcinated at 880-935$^{0}$C in air for
100 h with several intermediate grindings. The phase-purity of the
material was examined by powder x-ray diffraction. As synthesized,
the samples have oxygen contents in the range 6.975-6.985.
Starting with a material of maximal oxygen content a series of
reduced samples have been produced. Our characterization revealed
that the equilibration of the samples in closed ampoules with an
appropriate amount of copper powder reacting with oxygen leads to
the best results. Thus, for each sample in the series, an alumina
crucible with Y123 powder of exactly known weight and oxygen
content was placed in a quartz ampoule together with an exactly
weighted copper powder in an another crucible. To ensure a
homogenous oxygen distribution the ampoule was sealed under vacuum
and heated up to 700$^{0}$C (heating rate 10$^{0}$C/h), kept at
this temperature for 10h and finally slowly cooled (5$^{0} $C/h).
Field-cooled (FC) magnetization measurements were performed with a
Quantum Design SQUID magnetometer in a field of $0.5$~mT for
temperatures ranging from $5$~K to $100$~K. The Meissner fraction
$f$ was the deduced from the mass of the samples and the x-ray
density, and assuming spherical grains. To calculate the
temperature dependence of the effective penetration depth $\lambda
_{eff}$ $\left( T\right) $ we used the Shoenberg formula
\cite{shoenberg,porch} assuming spherical grains of radius $R$,
particle size distribution $N\left( R\right) $ and volume fraction
distribution $g\left( R\right) =N\left( R\right) R^{3}$,
\begin{equation}
f\left( T\right) =\int_{0}^{\infty }\left( 1-\frac{3\lambda \left(
T\right) }{R}\cot \left( \frac{R}{\lambda \left( T\right) }\right)
+\frac{3\lambda ^{3}\left( T\right) }{R^{2}}\right)
/\int_{0}^{\infty }g\left( R\right) dR \label{eq1a}
\end{equation}
We extracted the grain size distribution $N\left( R\right) $ from
an analysis of SEM (scanning electron microscope) photographs.
Solving this nonlinear equation for given $f\left( T\right) $ and
$g\left( R\right) $ we 0btained the effective penetration depth
$\lambda _{eff}$ $\left( T\right) $. For sufficiently anisotropic
superconductors ($\lambda _{c}/\lambda _{ab}>5 $), including
YBa$_{2}$Cu$_{3}$O$_{7-\delta }$ , $\lambda _{eff}$ is
proportional to the in-plane penetration depth, so that $\lambda
_{eff}=1.31\lambda _{ab}$\cite{fesenko}. The resulting data for
$\lambda _{ab}\left( T\right) $ agrees well with the
transverse-field $\mu $SR measurements performed on similar
samples \cite{Zimmermann95}.

\begin{figure}[tbp]
\centering
\includegraphics[totalheight=6cm]{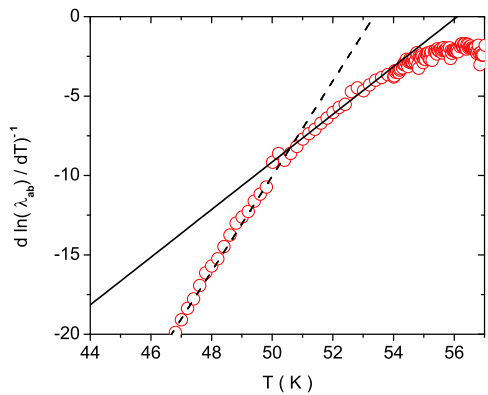}
\caption{$(d\ln \lambda _{ab}/dT)^{-1}$ with $\lambda _{ab}$ in
$\mu $m versus $T$ for YBa$_{2}$Cu$_{3}$O$_{6.59}$. The straight
line with slope $1/\nu \simeq 3/2$ corresponds according to Eq.
(\ref{eq1}) to charged criticality with $T_{c}=56.1$~K, while the
dashed line indicates the intermediate 3D-XY critical behavior
with slope $2/\nu \simeq 3$.} \label{fig1}
\end{figure}

When charged fluctuations dominate the in-plane penetration depth
and the correlation length are related
by\cite{herbut,herbut2,olsson,hove,mo}
\begin{equation}
\lambda _{ab}=\kappa _{ab}\xi _{ab},\text{ }\lambda _{ab}=\lambda
_{0ab}\left| t\right| ^{-\nu },\text{ }\nu \simeq 2/3,
\label{eq1}
\end{equation}
contrary to the uncharged case, where $\lambda \propto \sqrt{\xi
}$ and with that
\begin{equation}
\text{ }\lambda _{ab}=\lambda _{0ab}\left| t\right| ^{-\nu /2},
\label{eq2}
\end{equation}
where $t=T/T_{c}-1$. In a plot $(d\ln \lambda _{ab}/dT)^{-1}$
versus $T$ charged critical behavior is then uncovered in terms of
a temperature range where the data falls on a line with slope
$1/\nu \simeq 3/2$, while in the neutral (3D-XY) case it collapses
on a line with slope $2/\nu \simeq 3$. Clearly, in an
inhomogeneous system the phase transition is rounded and $(d\ln
\lambda _{ab}/dT)^{-1}$ does not vanish at $T_{c}$. In
Fig.~\ref{fig1} we displayed $(d\-ln \lambda _{ab}/dT)^{-1}$
versus $T$ for YBa$_{2}$Cu$_{3}$O$_{6.59}$, derived from the
measured in-plane penetration depth data. The data uncover a
crossover from uncharged critical behavior (dashed line) to
charged criticality (solid line) with $T_{c}\simeq 56.1$~K,
limited by a finite size effect due to the finite extent of the
grains and/or inhomogeneities within the grains. Although charged
criticality is attained there is no sharp transition. Indeed,
$\lambda _{ab}$ does not diverge at $T_{c}$ because the
correlation length $\xi _{ab}=\lambda _{ab}/\kappa _{ab}$ cannot
grow beyond the limiting length $L_{ab}$ in the $ab$-plane.

\begin{figure}[tbp]
\centering
\includegraphics[totalheight=6cm]{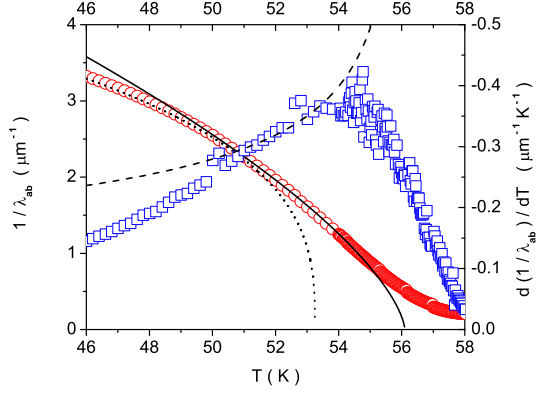}
\caption{$1/\lambda _{ab}$ and $d(1/\lambda _{ab})/dT$ \textit{vs}. $T$ for YBa%
$_{2}$Cu$_{3}$O$_{6.59}$. The solid line is $\lambda _{ab}=\lambda
_{0ab}\left| t\right| ^{-\nu }$ with $\lambda _{0ab}=0.089~\mu m$,
$\nu =2/3$ and $T_{c}=56.1$~K, appropriate for charged
criticality, and the dashed one its derivative. The dotted line
indicates uncharged critical behavior.} \label{fig2}
\end{figure}

To explore the evidence for charged critical behavior and the
nature of the finite size effect further, we displayed in Fig.
\ref{fig2} $1/\lambda _{ab}$ and $d(1/\lambda _{ab})/dT$
\textit{vs}. $T$. The solid line is $\lambda _{ab}=\lambda
_{0ab}\left| t\right| ^{-\nu }$ with $\lambda _{0ab}=0.089~\mu m
$, $\nu =2/3$ and $T_{c}=56.1$~K, appropriate for charged
criticality, and the dashed one its derivative. Approaching
$T_{c}$ of the fictitious homogeneous system the data reveals
again a crossover from uncharged to charged critical behavior,
while the tail in $\lambda _{ab}$ \textit{vs.} $T$ above $d\lambda
_{ab}/dT$ points to a finite size effect. Indeed, $d\lambda
_{ab}/dT$ \ does not diverge at $d\lambda _{ab}/dT$ but exhibits
an extremum at $T_{p}\simeq 54.27$~K, giving rise to an inflection
point in $\lambda _{ab}(T)$ at $T_{p}$. Here the correlation
length attains the limiting length.

\begin{figure}[tbp]
\centering
\includegraphics[totalheight=6cm]{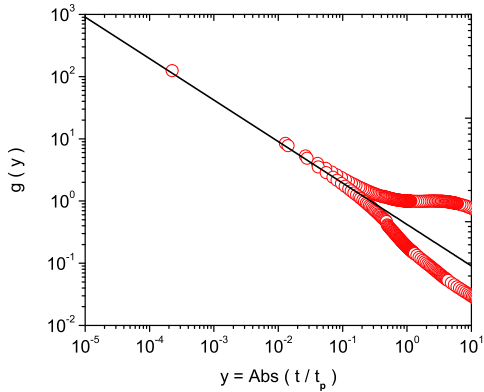}
\caption{Finite size scaling function $g\left( y\right) $ deduced
from the measured data with $\lambda _{0ab}=0.089~\mu m$, $\nu
=2/3$, $T_{c}=56.1$~K and $T_{p}\simeq 54.27$~K. The solid line
indicates the asymptotic behavior $g\left( y\rightarrow 0\right)
=g_{0}\left| y\right| ^{\nu }$ with $g_{0}=0.42 $.} \label{fig3}
\end{figure}

In this case the penetration depth adopts the finite size scaling
form \cite {cardy,schultka}
\begin{equation}
\lambda _{ab}\left( T\right) =\lambda _{0ab}\left| t\right| ^{-\nu
}g\left( y\right) ,\text{ \ }y=sign\left( t\right) \left|
\frac{t}{t_{p}}\right| \text{,}  \label{eq3}
\end{equation}
where $\xi _{ab}\left( T_{p}\right) =\xi _{0ab}\left| t_{p}\right|
^{-\nu }=L_{ab}$. For $t$ small and $L_{ab}\rightarrow \infty $
the scaling variable tends to $y\rightarrow \pm \infty $ where
$g\left( y\rightarrow -\infty \right) =1$ and $g\left(
y\rightarrow +\infty \right) =\infty $ while for $t=0$ and
$L_{ab}\neq 0$, $g\left( y\rightarrow 0\right) =g_{0}\left|
y\right| ^{\nu }=g_{0}\left| t/t_{p}\right| ^{\nu }$. In this
limit we obtain
\begin{equation}
\frac{\lambda _{ab}\left( T_{c},L_{ab}\right) }{\lambda
_{0ab}}=g_{0}\frac{L_{ab}}{\xi _{0ab}}.  \label{eq4}
\end{equation}
Another essential property of the finite size scaling function
stems from the existence of an inflection point in $\lambda
_{ab}\left( T\right) $. It implies an extremum in $d\lambda
_{ab}/dT$ at $T_{p}>T_{c}$ and with that the scaling form
$g^{+}\left( y\right) =y^{\nu }\left( 1+f\left( y\right) \right) $
with $df/dy\neq 0$ and $d^{2}f/dy^{2}=0$ at $y=1$, e.g. $f\left(
y\right) =ay+b\left( 1-y\right) ^{3}+c$. In Fig. \ref{fig3} we
displayed the finite size scaling function $g\left( y\right) $
deduced from the measured data with $\lambda _{0ab}=0.089\mu m$,
$\nu =2/3$, $T_{c}=56.1$~K and $T_{p}\simeq 54.27$~K. The solid
line indicates the asymptotic behavior $g\left( y\rightarrow
0\right) =g_{0}\left| y\right| ^{\nu }$ with $g_{0}=0.42 $. The
upper branch corresponding to $T<T_{c}$ tends to $g\left(
y\rightarrow -\infty \right) =1$, while the lower one referring to
$T>T_{c}$ approaches $g\left( y\rightarrow +\infty \right) =0$.
Consequently, the absence of a sharp transition (see Figs.~
\ref{fig1} and \ref{fig2}), is fully consistent with a finite size
effect arising from a limiting length $L_{ab}$ in the $ab$-plane,
attributable to the finite extent of the grains and/or
inhomogeneities within the grains.

\begin{figure}[tbp]
\centering
\includegraphics[totalheight=6cm]{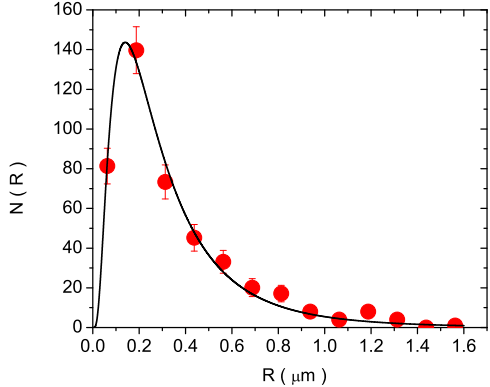}
\caption{Grain size distribution $N\left( R\right) $ of our
YBa$_{2}$Cu$_{3}$O$_{6.59}$ sample derived from an analysis of SEM
(scanning electron microscope) photographs. The solid line is a
fit to the log-normal distribution\protect\cite{kiefer}.}
\label{fig4}
\end{figure}

To disentangle these options we note that $L_{ab}/\xi _{0ab}\simeq
51$ follows from the finite size scaling analysis by invoking Eq.
(\ref{eq4}) with $\lambda _{ab}\left( T_{c},L_{ab}\right) $
$\simeq 1.898~\mu $m, $\lambda _{ab0}\left( T_{c},L_{ab}\right)
\simeq 0.089~\mu $m and $g_{0}=0.42$. Noting that $\xi
_{ab0}=\gamma \xi _{c0}$, where $\gamma $ is the anisotropy, we
obtain with $\gamma \approx 20$\cite{janossy} and $\xi
_{c0}\approx 10$~\AA, appropriate for
YBa$_{2}$Cu$_{3}$O$_{6.59}$\cite{tskulic}, the estimate
$L_{ab}\approx $ $510$~\AA\ compared to $L_{ab}\approx $
$572$~\AA\ found in YBa$_{2}$Cu$_{3}$O$_{6.7}$\cite{tsdc}. On the
other hand, a glance to Fig.\ref{fig4} shows that the grain size
distribution of our YBa$_{2}$Cu$_{3}$O$_{6.59}$ sample exhibits a
maximum at $2R=2800$~\AA\ and decreases steeply for smaller
grains. \ Hence, the smeared transition is not attributable to the
finite extent of the grains but due to inhomogeneities within the
grains. However, this does not point at bad sample quality but at
a genuine feature of underdoped cuprates reflecting the large
value of $\xi _{ab0}$ in this doping regime.

Nevertheless, our analysis of the in-plane penetration depth data
of underdoped YBa$_{2}$Cu$_{3}$O$_{6.59}$ provides remarkable
consistency for critical fluctuations, consistent with the charged
universality class, limited close to $T_{c}$ of the fictitious
infinite and homogeneous counterpart by a inhomogeneity induced
finite size effect. Since $\kappa _{ab}\propto \sqrt{T_{c}}$ this
will no longer hold true in the optimally doped counterparts. In
this doping regime there is mounting evidence for neutral (3D-XY)
critical behavior\cite {tshkws,book,kamal,parks,tskulic}. A
prominent example is YBa$_{2}$Cu$_{3}$O$_{6.95}$ revealing in the
in-plane penetration depth\cite{kamal} 3D-XY behavior over three
decades in reduced temperature,\ with no sign pointing to a
crossover to charged criticality.

In summary, we have presented in-plane penetration depth data for
YBa$_{2}$Cu$_{3}$O$_{6.59}$ providing consistent evidence for the
charged critical behavior of the superconductor to normal state
transition in type II superconductors ($\kappa >1/\sqrt{2}$).
Since the effective dimensionless charge $\widetilde{e}=\xi
/\lambda =1/\kappa $ scales as $T_{c}^{-1/2}$ this new critical
behavior should be observable and generic in suitably underdoped
cuprates. In this regime the crossover upon approaching $T_{c}$ is
thus to the charged critical regime, while near optimum doping it
is to the critical regime of a weakly charged superfluid where the
fluctuations of the order parameter are essentially those of an
uncharged superfluid (3D-XY). Furthermore, there is the
inhomogeneity induced finite size effect which renders the
asymptotic critical regime and with that the charged regime of
nearly optimally doped samples difficult to attain.

\acknowledgments The authors are grateful to D. Di Castro, I. F.
Herbut, S. Kohout, K.A. M\"{u}ller, J. Roos, A. Shengelaya and Z.
Tesanovic for very useful comments and suggestions on the subject
matter. This work was partially supported by the Swiss National
Science Foundation and the NCCR program \textit{Materials with
Novel Electronic Properties} (MaNEP) sponsored by the Swiss
National Science Foundation.

\end{document}